\documentclass[preprint,12pt]{elsarticle}

\usepackage{graphicx}
\usepackage{bm}
\usepackage{color}
\usepackage{gensymb}
\usepackage{a4wide}
\usepackage{hyperref}

\newcommand{\be}{\begin{equation}}
\newcommand{\ee}{\end{equation}}
\newcommand{\bea}{\begin{eqnarray}}
\newcommand{\eea}{\end{eqnarray}}

\newcommand{\ltsim}{\protect\raisebox{-0.5ex}{$\:\stackrel{\textstyle <}
	{\sim}\:$}}

\newcommand{\bvec}[1]{\ensuremath{\boldsymbol{#1}}}

\begin{document}

\begin{frontmatter}

  \title{Distillation of scalar exchange by coherent hypernucleus production in antiproton-nucleus collisions}

\author[lab1,lab2,lab3]{A.B. Larionov}
\ead{larionov@fias.uni-frankfurt.de}
\author[lab1]{H. Lenske}
\ead{Horst.Lenske@theo.physik.uni-giessen.de}

\address[lab1]{Institut f\"ur Theoretische Physik, Universit\"at Giessen,
               D-35392 Giessen, Germany}
\address[lab2]{National Research Center "Kurchatov Institute", 123182 Moscow, Russia}
\address[lab3]{Frankfurt Institute for Advanced Studies (FIAS),
               D-60438 Frankfurt am Main, Germany}

\date{\today}

\begin{abstract}
  The total and angular differential cross sections of the coherent process $\bar p + {}^AZ \to {}^A_\Lambda(Z-1) + \bar{\Lambda}$
  are evaluated at the beam momenta $1.5\div20$~GeV/c within the meson exchange model with bound proton and $\Lambda$-hyperon
  wave functions. It is shown that the shape of the beam momentum dependence of the hypernucleus production cross sections with various
  discrete $\Lambda$ states is strongly sensitive to the presence of the scalar $\kappa$-meson exchange in the $\bar p p \to \bar\Lambda \Lambda$
  amplitude. This can be used as a clean test of the exchange by scalar $\pi K$ correlation in coherent $\bar p A$ reactions.
\end{abstract}


\end{frontmatter}

\section{Introduction}
\label{intro}

Light scalar mesons represent one of the most puzzling areas in the quark/hadron physics as their structure
is different from $q \bar q$. In particular, identifying the $\kappa$ ($K_0^*(800)$) \cite{Olive:2016xmw}
in various hadronic processes is of importance since this meson is a candidate member of the hypothetical
SU(3) octet of scalar mesons with masses below 1 GeV.
Apart from $\bar\kappa$, the octet-partners of $\kappa$ are the non-strange isoscalar $\sigma$ ($f_0(500)$)
and isovector $\delta$ ($a_0(980)$) mesons.
So far the corresponding particles are only seen as broad resonance-like structures in the $0^+$-meson spectrum.
The $\sigma$, $\delta$, $\kappa$ and $\bar\kappa$ are likely the $\pi\pi$, $\pi\eta$, $\pi K$ and $\pi \bar K$
resonance states, respectively.
In that sense, the $\kappa$ exchange can be regarded as an economical way to take into account the correlated $\pi K$ channel.
The $\kappa$ exchange channel is of particular interest for multi-strangeness
baryonic matter in heavy ion collisions and in neutron stars.
This channel is also an indispensable part of baryon-baryon interaction approaches
utilizing the SU(3) flavour \cite{Timmermans:1992fu} and SU(6) spin-flavour \cite{Haidenbauer:2005zh} group structures.

The purpose of this letter is to study $\Lambda$-hypernucleus production in antiproton-nucleus reactions initiated by the
$\bar p p \to \bar\Lambda \Lambda$ process on the bound proton. The produced $\Lambda$ is captured to one of the shells
of the outgoing hypernucleus. The main emphasis is put on the influence of the inclusion of the $\kappa$-meson on
the reaction cross section.

\begin{figure}
\begin{center}
\includegraphics[scale = 0.6]{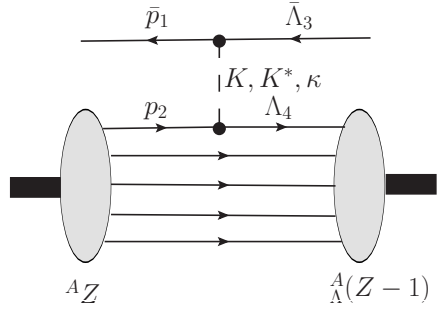}
\caption{\label{fig:hyperProd} The Feynman graph of the ${}^AZ(\bar p,\bar\Lambda){}^A_\Lambda(Z-1)$ process.
  Dashed line represents the propagator of the exchange meson.
  The grey ellipsoids correspond to the wave functions
  of the initial ground state nucleus $^AZ$ and final hypernucleus \mbox{$^A_\Lambda(Z-1)$}.}
\end{center}
\end{figure}

The full process is sketched in Fig.~\ref{fig:hyperProd}.
At momentum transfers $\ltsim 1$ GeV/c, the amplitude of the $\bar p p \to \bar\Lambda \Lambda$ reaction can be described by $t$-channel
exchanges of the light mesons with strangeness $|S|=1$, such as pseudoscalar $K$, vector $K^*$, and, probably, hypothetical
scalar $\kappa$ mesons, respectively. If the initial proton and final $\Lambda$ are bound, the unnatural parity $K$-exchange is strongly suppressed,
since it is pure relativistic effect proceeding through the lower wave function components of either proton
or $\Lambda$ Dirac spinors. Hence the coherent hypernucleus production should filter-out the kaon exchange
allowing to exclusively address the exchange of the natural parity $K^*$ and $\kappa$. As it will be shown, the scalar $\kappa$-exchange
contribution to the hypernucleus production cross section is very well visible on the background of smoothly growing with beam momentum
vector $K^*$-exchange contribution. Therefore the nuclear binding helps to effectively distillate the $\kappa$ meson from its mixture
with kaon exchange in free-space $\bar p p \to \bar\Lambda \Lambda$ amplitude.

The paper is structured as follows. In sec. \ref{model} we describe the strange meson exchange model starting from the basic Lagrangians.
The amplitude of the hypernucleus production is derived in the impulse approximation
(IA) and afterwards corrected for the initial and final state interactions (ISI/FSI) of $\bar p$ and $\bar\Lambda$ in the nucleus.
We introduce two sets of model parameters, both of which describe the momentum dependence
of the total $\bar p p \to \bar\Lambda \Lambda$ cross section. The first set does not include $\kappa$-meson exchange
and the second set includes it. The wave functions of the bound proton and $\Lambda$ states are obtained by solving the Dirac equation
with relativistic mean fields (RMF).
Section \ref{results} contains numerical results. The angular differential and total cross sections of the hypernucleus
production with $\Lambda$ on various shells are calculated. We emphasize that the shape of the momentum dependence of total
cross sections is qualitatively different in calculations with and without $\kappa$. The summary and outlook
are given in sec. \ref{SumConcl}.

\section{The Model}
\label{model}

We will introduce the $K$, $K^*$ and $\kappa$ exchanges by using the following
interaction Lagrangians \cite{Cheoun:1996kn,Tsushima:1998jz,Han:1999ck}:
\begin{eqnarray}
   {\cal L}_{KN\Lambda}  &=& -ig_{KN\Lambda} \bar N \gamma^5 \Lambda K + \mbox{h.c.}~,     \label{Lag_KNL}\\
   {\cal L}_{K^*N\Lambda} &=& \bar N \left(G_v\gamma^\mu - \frac{G_t\sigma^{\mu\nu}\partial_\nu^{K^*}}{m_N+m_\Lambda}\right)\Lambda K^*_\mu
                           + \mbox{h.c.}~,      \label{Lag_KsNL}\\
   {\cal L}_{\kappa N\Lambda} &=& -g_{\kappa N\Lambda} \bar N \Lambda \kappa  + \mbox{h.c.}~.  \label{Lag_kappaNL}
\end{eqnarray}
In the case of the bound proton and $\Lambda$ we include their wave functions in the field operators of the Lagrangians
Eqs.~(\ref{Lag_KNL})-(\ref{Lag_kappaNL}) and calculate the $S$-matrix in the second order perturbation theory
using Wick's theorem (cf. Ref. \cite{BLP}):
\begin{equation}
   S=\frac{2\pi\delta(E_1+E_2-E_3-E_4)}{(2E_1V 2E_3V)^{1/2}} i({\cal M}_K + {\cal M}_{K^*} + {\cal M}_{\kappa})~,       \label{S}
\end{equation}
where $E_i,~i=1,2,3,4$ are particle energies (see Fig.~\ref{fig:hyperProd} for notation) and $V$ is the normalization volume.
The $K,~K^*$ and $\kappa$ exchange (noninvariant) partial matrix elements are expressed as
\begin{eqnarray}
  {\cal M}_K &=& -g^2_{KN\Lambda} F^2_K(t) \sqrt{\Omega}\,\bar u_{-p_1,-\lambda_1} \gamma^5 u_{-p_3,-\lambda_3}
  \frac{1}{t-m_K^2}
  \int d^3r e^{-i\bm{q}\bm{r}} \bar \psi_4(\bm{r}) \gamma^5 \psi_2(\bm{r})~,            \label{calM_K}\\
  {\cal M}_{K^*} &=& -F^2_{K^*}(t) \sqrt{\Omega}\,\bar u_{-p_1,-\lambda_1} \Gamma^\mu(-q) u_{-p_3,-\lambda_3}
  G_{\mu\nu}(q)
  \int d^3r e^{-i\bm{q}\bm{r}} \bar \psi_4(\bm{r}) \Gamma^\nu(q) \psi_2(\bm{r})~,        \label{calM_Ks}\\
  {\cal M}_{\kappa} &=& g^2_{\kappa N\Lambda} F^2_{\kappa}(t) \sqrt{\Omega}\,\bar u_{-p_1,-\lambda_1} u_{-p_3,-\lambda_3}
  \frac{1}{t-m_{\kappa}^2}
  \int d^3r e^{-i\bm{q}\bm{r}} \bar \psi_4(\bm{r}) \psi_2(\bm{r})~.                     \label{calM_kappa}
\end{eqnarray}
Here, $\psi_2(\bm{r})$ and $\psi_4(\bm{r})$ are the relativistic wave functions of the bound proton and $\Lambda$, respectively.
They satisfy the normalization conditions:
\begin{equation}
   \int d^3r \psi_i^\dag(\bm{r}) \psi_i(\bm{r}) = 1~,~~~i=2,4~.                      \label{normCond}
\end{equation}
$p_i$ is the four-momentum and $\lambda_i=\pm1/2$ is the spin projection of an antiproton ($i=1$) and antilambda ($i=3$).
$q=p_3-p_1$ is the four-momentum transfer, $t=q^2$. In Eq.~(\ref{calM_Ks}),
\begin{equation}
   G_{\mu\nu}(q) = \frac{-g_{\mu\nu} + q_\mu q_\nu/m_{K^*}^2}{t-m_{K^*}^2}     \label{G_mu_nu}
\end{equation}
is the $K^*$ meson propagator. We neglected the widths of the $K^*$ and $\kappa$ mesons in their propagators
as the momentum transfers are space-like (e.g. $-t=0.08 \div 1.7$ GeV$^2$ at $p_{\rm lab}=2$ GeV/c).
The $K^*N\Lambda$ vertex function is defined as
\begin{equation}
   \Gamma^\mu(q)=iG_v\gamma^\mu + \frac{G_t}{m_N+m_\Lambda} \sigma^{\mu\nu} q_\nu~.    \label{Gamma^mu}
\end{equation}
The vertex form factors are chosen in the monopole form:
\begin{equation}
   F_j(t) = \frac{\Lambda_j^2-m_j^2}{\Lambda_j^2-t}~,~~~j=K,K^*,\kappa~.       \label{FFs}
\end{equation}
Similar to Refs. \cite{Sopkovich,Shyam:2014dia,Shyam:2015hqa} we included in Eqs.~(\ref{calM_K})-(\ref{calM_kappa}) the attenuation
factor $\sqrt{\Omega}$ to describe the modification of the elementary $\bar p p \to \bar \Lambda \Lambda$ amplitude due to ISI in the $\bar p p$ channel
and FSI in the $\bar \Lambda \Lambda$ channel. This effectively takes into account the absorptive $\bar p p$ and $\bar \Lambda \Lambda$ potentials.
For simplicity, we assume the attenuation factor $\Omega$ to be energy independent.
With $\Omega=1$, Eqs.~(\ref{calM_K})-(\ref{calM_kappa}) correspond to the Born approximation.
The Dirac spinors of the antiproton and antilambda plane waves are normalized according to Ref. \cite{BLP}:
$\bar u_{-p,-\lambda} u_{-p,-\lambda} = -2m_{N(\Lambda)}$.

The differential cross section in the rest frame of the target nucleus is written as
\begin{equation}
  d\sigma = \frac{2\pi\delta^{(4)}(p_1+p_A-p_3-p_B)}{2p_{\rm lab}}\,
           \frac{1}{2} \sum_{\lambda_1,m_2,\lambda_3,m_4} |{\cal M}_K + {\cal M}_{K^*} + {\cal M}_{\kappa}|^2   \,
            \frac{d^3p_3}{(2\pi)^32E_3} d^3p_B~,                 \label{dSigma}
\end{equation}
where $p_A$ and $p_B$ are the four-momenta of the initial nucleus ($A$) and final hypernucleus ($B$).
The $\delta$ function in Eq.~(\ref{dSigma}) takes into account the recoil of the hypernucleus.
The cross section is summed over the total angular momentum projections $m_2$ and $m_4$ of the bound target proton
and of the $\Lambda$ hyperon, respectively, and over the spin projection $\lambda_3$ of the outgoing $\bar\Lambda$.
The averaging is taken over the spin projection $\lambda_1$ of the incoming antiproton, as expressed by the factor 1/2.

The choice of coupling constants is based on SU(3) relations \cite{deSwart:1963pdg}:
\begin{eqnarray}
   g_{KN\Lambda}      &=& -g_{\pi NN} \frac{3-2\alpha_{PS}}{\sqrt{3}}~,      \label{g_KNL}\\
   G_{v,t}           &=& -G_{v,t}^\rho \frac{3-2\alpha_{E,M}}{\sqrt{3}}~,    \label{G_vt}\\
   g_{\kappa N\Lambda}  &=& -g_{\sigma NN} \frac{3-2\alpha_S}{3-4\alpha_S}~,    \label{g_kappaNL}
\end{eqnarray}
where $\alpha$'s are the $D$-type coupling ratios.
The $\pi NN$ coupling constant is very well known, $g_{\pi NN}=13.4$  \cite{Dumbrajs:1983jd}.
The vector $\rho NN$ coupling constant is also fixed, $G_{v}^\rho=2.66$, however, the tensor $\rho NN$ coupling constant
is quite uncertain, $G_{t}^\rho=10.9\div20.6$ \cite{Cheoun:1996kn}.
The $\sigma NN$ coupling constant can be estimated either
from the Bonn model \cite{Machleidt:1987hj} or from the Walecka-type models (cf. \cite{Lalazissis:1996rd}).
In both cases one obtains $g_{\sigma NN} \simeq 10$.
The $\alpha$'s for the octets of light pseudoscalar and vector mesons
are reasonably well determined \cite{Cheoun:1996kn,Han:1999ck}:
$\alpha_{PS} \simeq 0.6$, $\alpha_{E} \simeq 0$, $\alpha_{M} \simeq 3/4$. However, safe
phenomenological information on $\alpha_S$ is lacking.

Thus, the coupling constants $G_{t}$ and $g_{\kappa N\Lambda}$,
the cutoff parameters $\Lambda_K, \Lambda_{K^*}$ and $\Lambda_\kappa$, and the attenuation factor $\Omega$
remain to be determined from comparison with experimental data. We adjusted these parameters to describe
the beam momentum dependence of the $\bar p p \to \bar\Lambda \Lambda$ cross section. The two sets
of parameters, (1) without $\kappa$ meson and (2) with $\kappa$ meson, are listed in Table~\ref{tab:par}.
In the calculations we used the mass $m_\kappa=682$ MeV \cite{Olive:2016xmw}.
\begin{table}[htb]
  \caption{\label{tab:par}
    Parameters of the $\bar p p \to \bar\Lambda \Lambda$ amplitude.
    The value of $g_{KN\Lambda}$ slightly differs from
    -13.3 as given by Eq.~(\ref{g_KNL}) and is taken from
    $K^+N$ scattering analysis of Ref. \cite{Buettgen:1990yw}.
    The cutoff parameters $\Lambda_K$, $\Lambda_{K^*}$ and $\Lambda_\kappa$ are in GeV. The attenuation factors are shown in the last column. }
  \begin{center}
    \begin{tabular}{|c|c|c|c|c|c|c|c|c|}
    \hline
    Set~~~& $g_{KN\Lambda}$~~~~~& $G_{v}$~~~~~& $G_{t}$~~~~~& $g_{\kappa N\Lambda}$~~~~~& $\Lambda_K$~~~~~& $\Lambda_{K^*}$~~~~~&\
    $\Lambda_\kappa$~~~~~& $\Omega$~~~~~\\
    \hline
    1  & -13.981       &  -4.6     & -8.5     &  ---               &   2.0         &  1.6            & ---             & 0.0150 \\
    2  & -13.981       &  -4.6     & -9.0     &  -7.5              &   1.8         &  2.0            & 1.8             & 0.0044 \\
    \hline
  \end{tabular}
  \end{center}
\end{table}

\begin{figure}
\begin{center}
   \includegraphics[scale = 0.4]{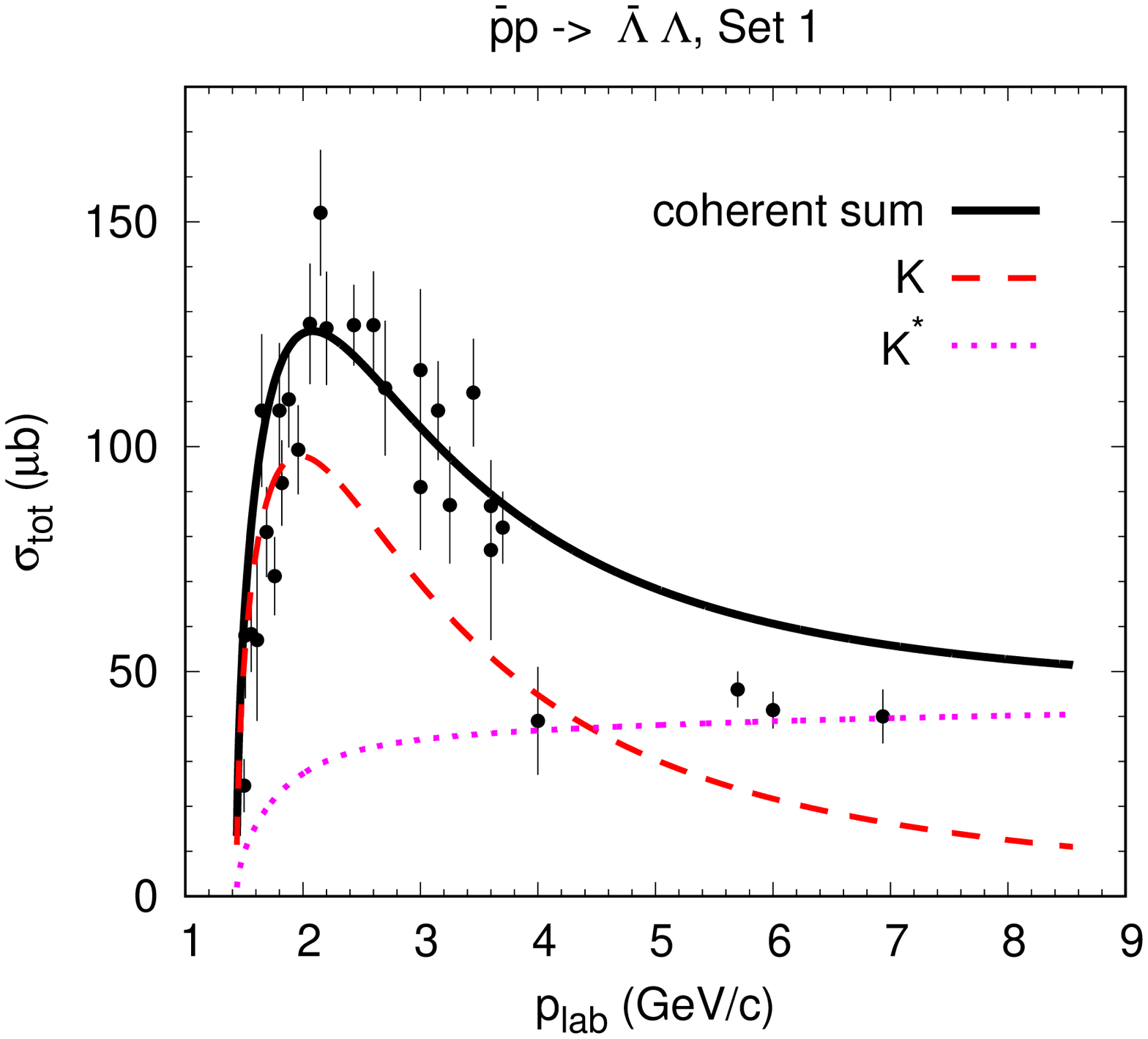}
   \includegraphics[scale = 0.4]{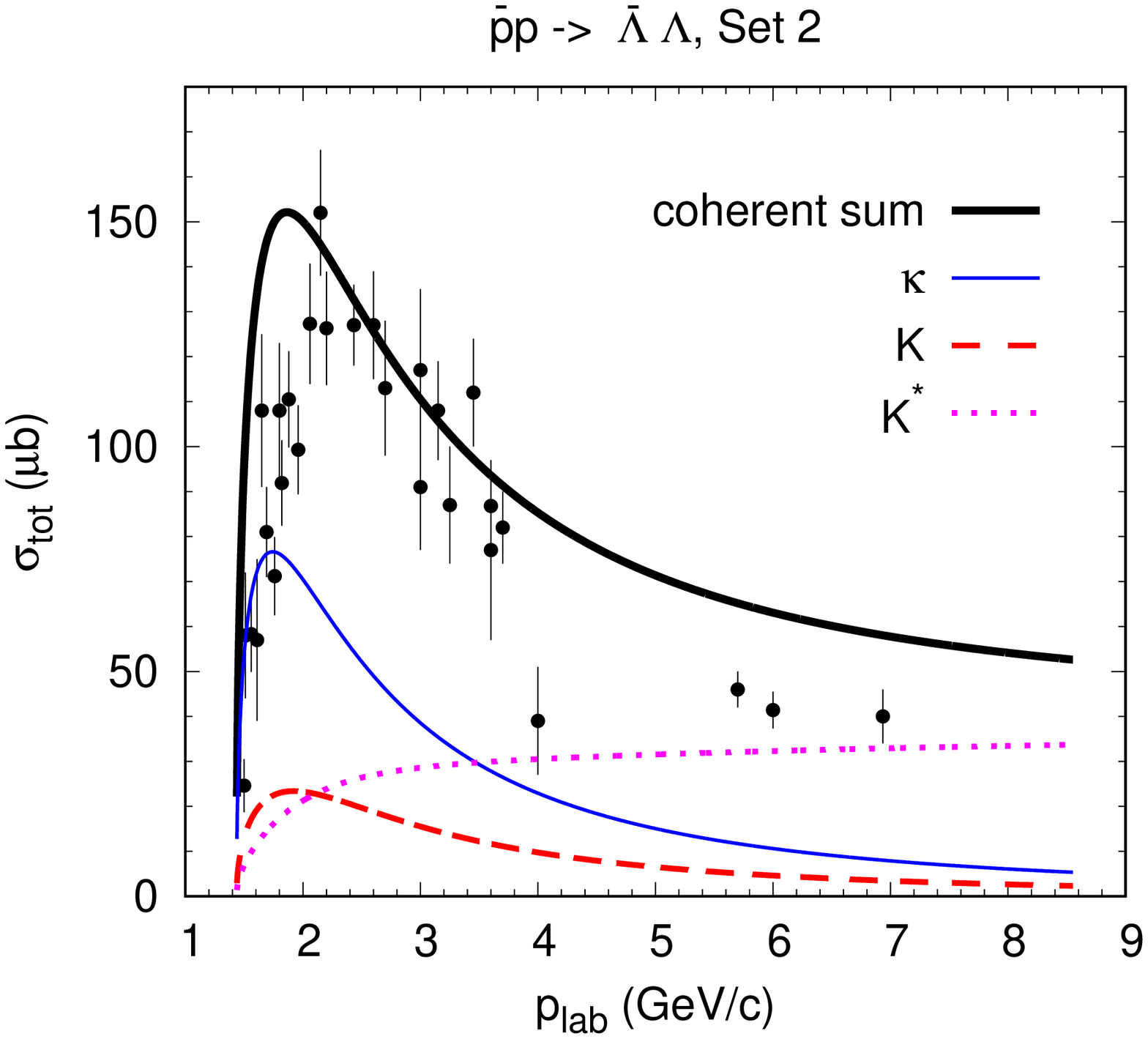}
\end{center}
\caption{\label{fig:sigma_Lbar_L} Angle-integrated cross section of the process $\bar p p \to \bar\Lambda \Lambda$
  as a function of the beam momentum calculated without (Set 1) and with (Set 2) inclusion of the $\kappa$ meson.
  Full calculations are shown by thick solid lines. Other lines, as indicated, show the partial contributions
  of the $K$, $K^*$ and $\kappa$ exchanges. 
  Experimental data are from Ref. \cite{Bald87}.}
\end{figure}
As we see from Fig.~\ref{fig:sigma_Lbar_L}, in the calculation with set 1 the peak of the $\bar p p \to \bar\Lambda \Lambda$
cross section at $p_{\rm lab} \simeq 2$ GeV/c is saturated by the $K$ exchange.
In contrast, in the case of set 2 the peak is saturated mostly by the $\kappa$ exchange. The $K^*$-exchange contribution grows monotonically
with beam momentum and becomes dominant at $p_{\rm lab} > 3\div4$ GeV/c. We note that set 2 gives steeper increasing angular differential cross
section towards $\Theta_{\rm c.m.}=0$ at $p_{\rm lab}=2.060$ GeV/c in a better agreement with experimental data \cite{Jayet:1978yq} than set 1.

The matrix elements Eqs.~(\ref{calM_K})-(\ref{calM_kappa}) are obtained in the impulse approximation (IA).
More realistic calculation should take into account the distortion of the incoming $\bar p$ and outgoing $\bar\Lambda$
waves, mostly due to strong absorption of the antibaryons in the nucleus. In the eikonal approximation the incoming
$\bar p$ wave is multiplied by the factor
\begin{equation}
  F_{\bar p}(\bm{r}) =
  \exp\left(-\frac{1}{2}\sigma_{\bar pN}(1-i\alpha_{\bar pN}) \int\limits_{-\infty}^0 d\xi
        \rho(\bm{r}+\frac{\bm{p}_{\bar p}}{p_{\bar p}}\xi)\right)~,   \label{F_barp}
\end{equation}
and the outgoing $\bar\Lambda$ wave is multiplied by
\begin{equation}
  F_{\bar\Lambda}(\bm{r}) =
  \exp\left(-\frac{1}{2}\sigma_{\bar \Lambda N}(1-i\alpha_{\bar \Lambda N}) \int\limits_0^{+\infty} d\xi
        \rho(\bm{r}+\frac{\bm{p}_{\bar\Lambda}}{p_{\bar\Lambda}}\xi)\right)~,
                     \label{F_barL}
\end{equation}
where $\rho(\bm{r})$ is the nucleon density, $\sigma_{jN}$ is the total $jN$ cross section,
$\alpha_{jN}=\mbox{Re}f_{jN}(0)/\mbox{Im}f_{jN}(0)$ is the ratio of the real-to-imaginary part of the forward $jN$ amplitude
($j=\bar p, \bar\Lambda$).
Equations (\ref{F_barp}),(\ref{F_barL}) can be obtained by applying the eikonal approximation to solve the Schr\"odinger equation
for the scattering of a particle in the external potential (cf. Ref. \cite{LL}) which is then replaced by the optical potential
in the low-density approximation. The integrands in the matrix elements Eqs.~(\ref{calM_K})-(\ref{calM_kappa}) are then multiplied
by $F_{\bar p}(\bm{r}) F_{\bar\Lambda}(\bm{r})$ (cf. Refs. \cite{Bando:1990yi,Frankfurt:1994nn}).
In numerical calculations we applied the momentum dependent total $\bar pN$ cross
section and the ratio $\alpha_{\bar p N}$ as described in Ref. \cite{Larionov:2016xeb}. We assumed that
$\sigma_{\bar\Lambda N} = \sigma_{\bar p N}$ at the same beam momenta which is supported by experimental data on the total $\bar\Lambda p$
cross section at $p_{\rm lab}=4\div14$ GeV/c \cite{Eisele:1976fe}. For simplicity we have set $\alpha_{\bar \Lambda N}=0$.

The nucleon and hyperon ($B=N,\Lambda$) single particle bound state wave functions are determined
as solutions of a static Dirac equation with scalar and vector potentials (cf. Refs. \cite{Bender:2009cj,Glendening:1992du,Keil:1999hk}):
\begin{equation}\label{statDirac}
\left( -i \bvec{\alpha}\cdot \bvec{\nabla} + \beta m^*_B(r) + V_B(r)+V_C(r)- \varepsilon \right) \psi_B(\bm{r}) =0~,
\end{equation}
where $m^*_B(r)=m_B+S_B(r)$ is the effective (Dirac) mass.
Both the scalar ($S_B$) and nuclear vector ($V_B$) potentials are chosen in the form of superpositions of
the classical meson fields, $\sigma(I=0,J^{P}=0^+),~\omega(0,1^-),~\delta(1,0^+)$ and $\rho(1,1^-)$, weighted by the strong interaction
coupling constants appropriate for the given baryon. The meson fields are parameterized by Woods-Saxon form factors.
For protons also the static Coulomb potential ($V_C$) contributes \cite{Keil:1999hk}.
We will consider the case of a spherical nucleus. Hence the eigenfunctions of the Dirac equation are characterized by
radial, orbital and total angular momentum quantum numbers,
$n,~l,~j$, respectively, and the magnetic quantum number $m \equiv j_z$.

In the future \={P}ANDA experiment at FAIR the noble gases will be used as targets.
Thus as a representative case we consider the reaction
$
   \bar p + {}^{40}\mbox{Ar} \to {}^{40}_{~\Lambda}\mbox{Cl} +  \bar\Lambda
$.
Using the RMF approach the parameters of the Woods-Saxon form factors have been chosen to fit
the binding energy, single-particle separation energies, and the root-mean-square radii of the nucleon density distributions
in the ${}^{40}\mbox{Ar}$ nucleus. Under the assumption that the nuclear potentials do not change after a sudden removal of the valence proton,
the $\Lambda$-hyperon scalar and vector potentials in the ${}^{40}_{~\Lambda}\mbox{Cl}$ hypernucleus were obtained by multiplying
the scalar and vector nucleon potentials in the ${}^{40}\mbox{Ar}$ nucleus by the factors
$0.525$ and $0.550$, respectively. This leads to a good agreement of the $\Lambda$ energy levels with the empirical
systematics and with the previous relativistic mean-field calculations \cite{Keil:1999hk}.
In order to assure that after the reaction the residual core nucleus carries as little excitation energy as possible,
we consider only strangeness creation processes on protons of the $1d_{3/2}$ valence shell in ${}^{40}\mbox{Ar}$.

\section{Numerical results}
\label{results}

\begin{figure}
\begin{center}
  \includegraphics[scale = 0.6]{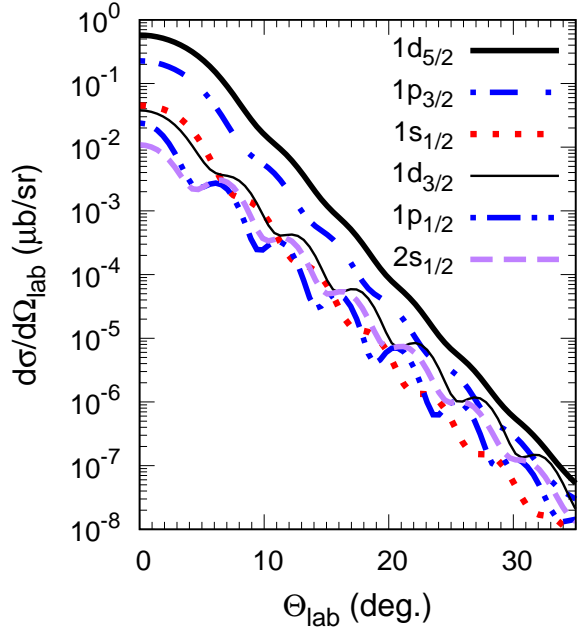}
\end{center}
\caption{\label{fig:dsigdO_sum} Angular differential cross section of the reaction ${}^{40}\mbox{Ar}(\bar p,\bar \Lambda){}^{40}_{~\Lambda}\mbox{Cl}$
  at $p_{\rm lab}=2$ GeV/c. Lines show the calculations for $\Lambda$ in various states, as indicated. Calculations include $\kappa$ exchange.}
\end{figure}
\begin{figure}
\begin{center}
   \includegraphics[scale = 0.5]{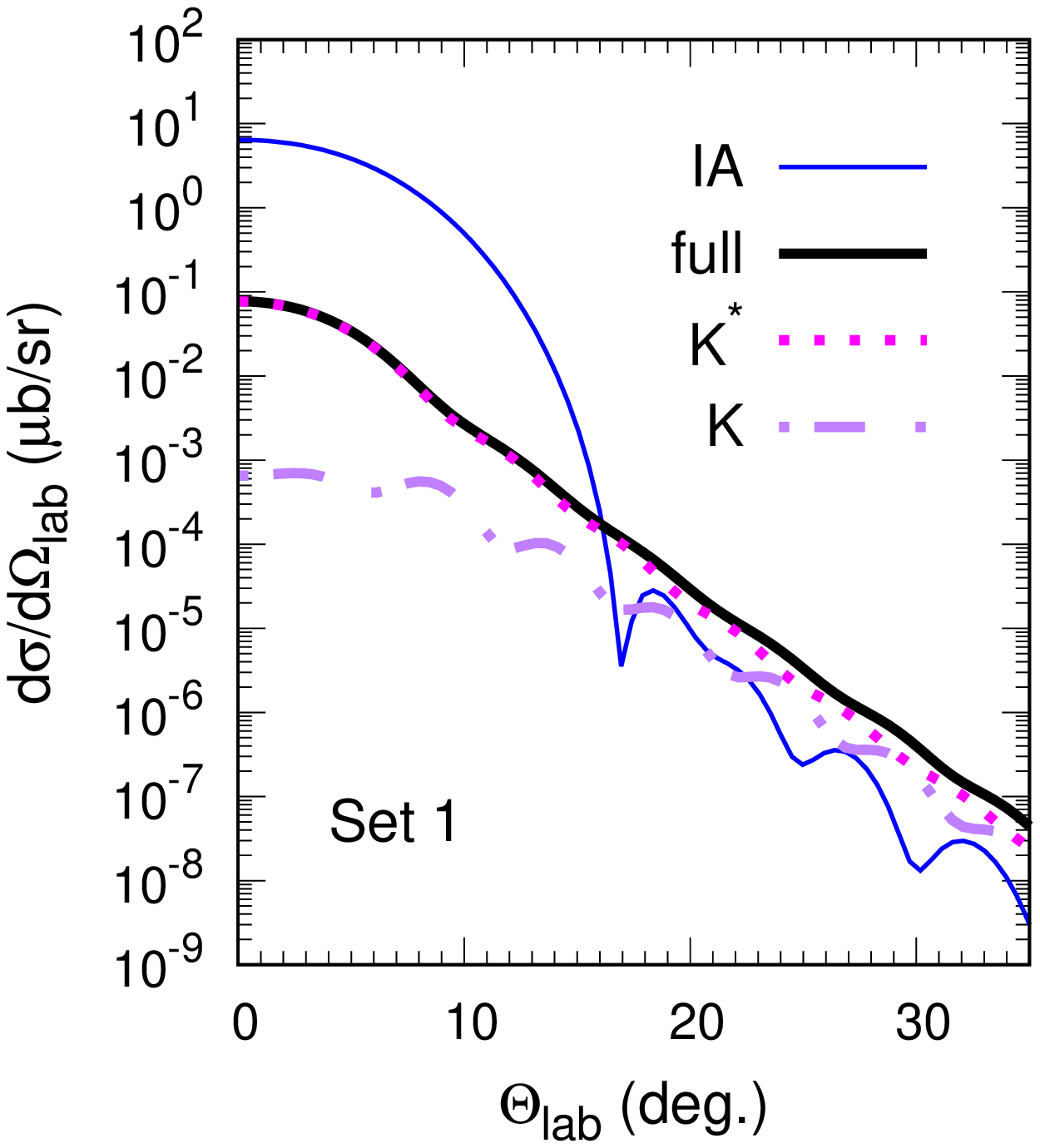}
   \includegraphics[scale = 0.5]{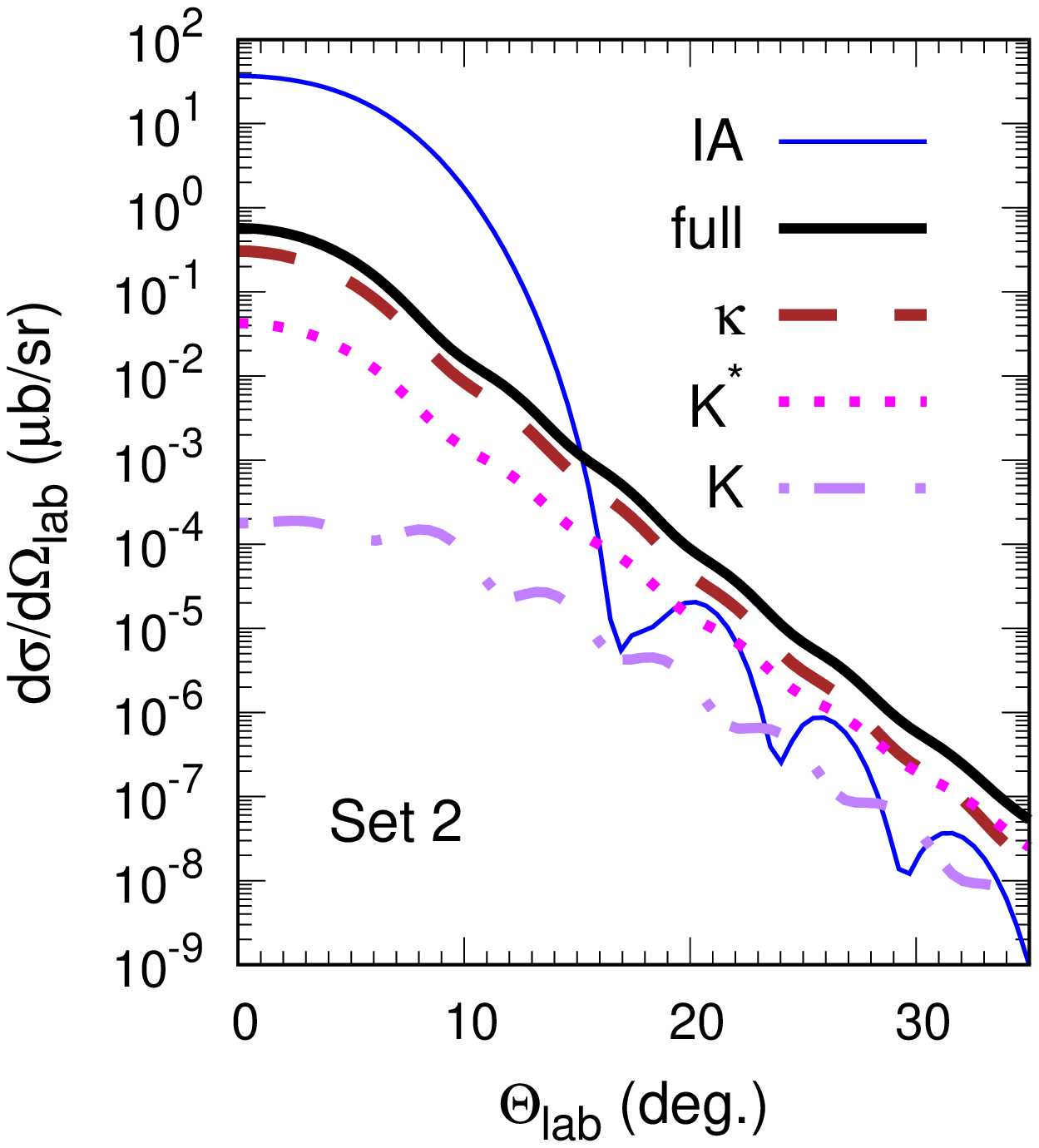}
\end{center}
\caption{\label{fig:dsigdO_1d2.5} Angular differential cross section of the reaction ${}^{40}\mbox{Ar}(\bar p,\bar \Lambda){}^{40}_{~\Lambda}\mbox{Cl}$
  at $p_{\rm lab}=2$ GeV/c with $1d_{5/2}$ $\Lambda$ state.
  The IA calculation, the full calculation (with absorption), and the separate meson contributions
  to the full calculation are shown by different lines.
  Results without (Set 1, left panel) and with (Set 2, right panel) $\kappa$ meson are displayed as indicated.}
\end{figure}
The differential hypernuclear production cross sections with the $\Lambda$ occupying various shells are displayed in Fig.~\ref{fig:dsigdO_sum}.
Irrespective of spin-orbit effects
\footnote{There is a strong difference between the cross sections with $j_\Lambda = l_\Lambda \pm 1/2$, so that the cross sections for the larger value
  of $j_\Lambda$ are much larger at forward angles. This effect arises from the detailed structure of the Fourier transforms of the transition form factors
  in the matrix elements Eqs.~(\ref{calM_K})-(\ref{calM_kappa}) and its study is beyond the scope of this work.},
the cross sections are larger for larger hyperon orbital angular momentum, $l_\Lambda$.
We have checked that within the IA the cross sections at $\Theta_{\rm lab}=0$  for the $1d_{5/2}$, $1p_{3/2}$ and $1s_{1/2}$ hyperon states are nearly equal.
Thus the rise of the cross sections at forward laboratory angles with $l_\Lambda$ is mostly caused by the nuclear absorption which is diminished with increasing
$l_\Lambda$ due to the shift of the hyperon density distribution to larger radii.
The largest cross section is obtained for the ${}^{40}_{~\Lambda}\mbox{Cl}$ hypernucleus with $\Lambda$ in the $1d_{5/2}$ state.
The differential angular distribution for this case is analyzed in more detail in Fig.~\ref{fig:dsigdO_1d2.5}.
From the comparison of the full and IA calculations we observe that the absorption of $\bar p$ and $\bar \Lambda$ has a quite significant effect:
it reduces the cross section drastically, amounting at forward angles
to about two orders of magnitude, and smears out the diffractive structures.
Similar effects of the absorption are present also for the other $\Lambda$ states.

A deeper insight into the production mechanism is obtained by decomposing the total reaction amplitude into different meson exchange parts.
From the partial meson exchange contributions, shown in Fig.~\ref{fig:dsigdO_1d2.5}, it is remarkable that for Set 1 the kaon contribution is small
and the spectrum is dominated by $K^*$, however, from Fig.~\ref{fig:sigma_Lbar_L} one would expect the opposite.
This surprising result can be understood by the fact that the momentum transfer to the $\bar\Lambda$ is provided by the nucleus as a whole while
the produced $\Lambda$ is almost at rest. The exchange by pseudoscalar meson is suppressed in this case since it proceeds through the lower components
of the proton and hyperon Dirac spinors\footnote{Formally this is due to the $\gamma^5$ matrix in the integrand of Eq.~(\ref{calM_K}).}
which are suppressed by factors $\sim 1/m_BR$, where $R$ is the nuclear radius.
In contrast, in the case of the free space $\bar p p \to \bar\Lambda \Lambda$ process
the $\Lambda$ is produced with finite momentum. Therefore the upper and lower components of its Dirac spinor
are of comparable magnitude which favours the pseudoscalar meson exchange.

The situation is very different in the case of Set 2. Here, $\kappa$ plays the dominant role both for the free scattering
$\bar p p \to \bar\Lambda \Lambda$ and for the hypernucleus production since the scalar exchange is not suppressed in the recoilless kinematics.

\begin{figure}
\begin{center}
   \includegraphics[scale = 0.5]{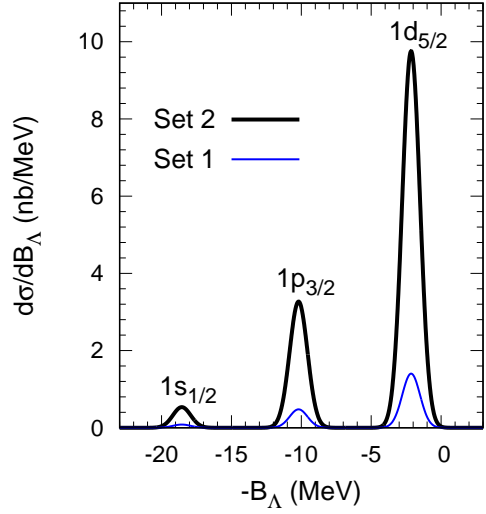}
\end{center}
\caption{\label{fig:Ar40_spectrum} Spectral distribution of $\Lambda$ bound states in a ${}^{40}_{~\Lambda}\mbox{Cl}$
hypernucleus produced in coherent $\bar p\,{}^{40}\mbox{Ar}$ collisions at $p_{\rm lab}=2$ GeV/c. The angle-integrated cross sections for the hypernucleus production
  in $1s_{1/2}$, $1p_{3/2}$ and $1d_{5/2}$ states have been folded by Gaussians of width $\Gamma_{FWHM}=1.5$~MeV which is a typical experimental energy resolution. Thick and thin solid lines correspond to the results obtained with (Set 2) and without (Set 1) $\kappa$ exchange, respectively.}
\end{figure}
As we see from Fig.~\ref{fig:Ar40_spectrum}, the cross section of coherent hypernucleus production in different states is much larger
when the $\kappa$ exchange is included. This is pure quantum coherence effect since the angle-integrated
$\bar p p \to \bar\Lambda \Lambda$ cross sections differ by $\sim 15\%$ only at $p_{\rm lab}=2$ GeV/c
(Fig.~\ref{fig:sigma_Lbar_L}) while the hypernuclear production cross sections differ by almost one order of magnitude for Set 1 and Set 2.

\begin{figure}
\begin{center}
     \includegraphics[scale = 0.5]{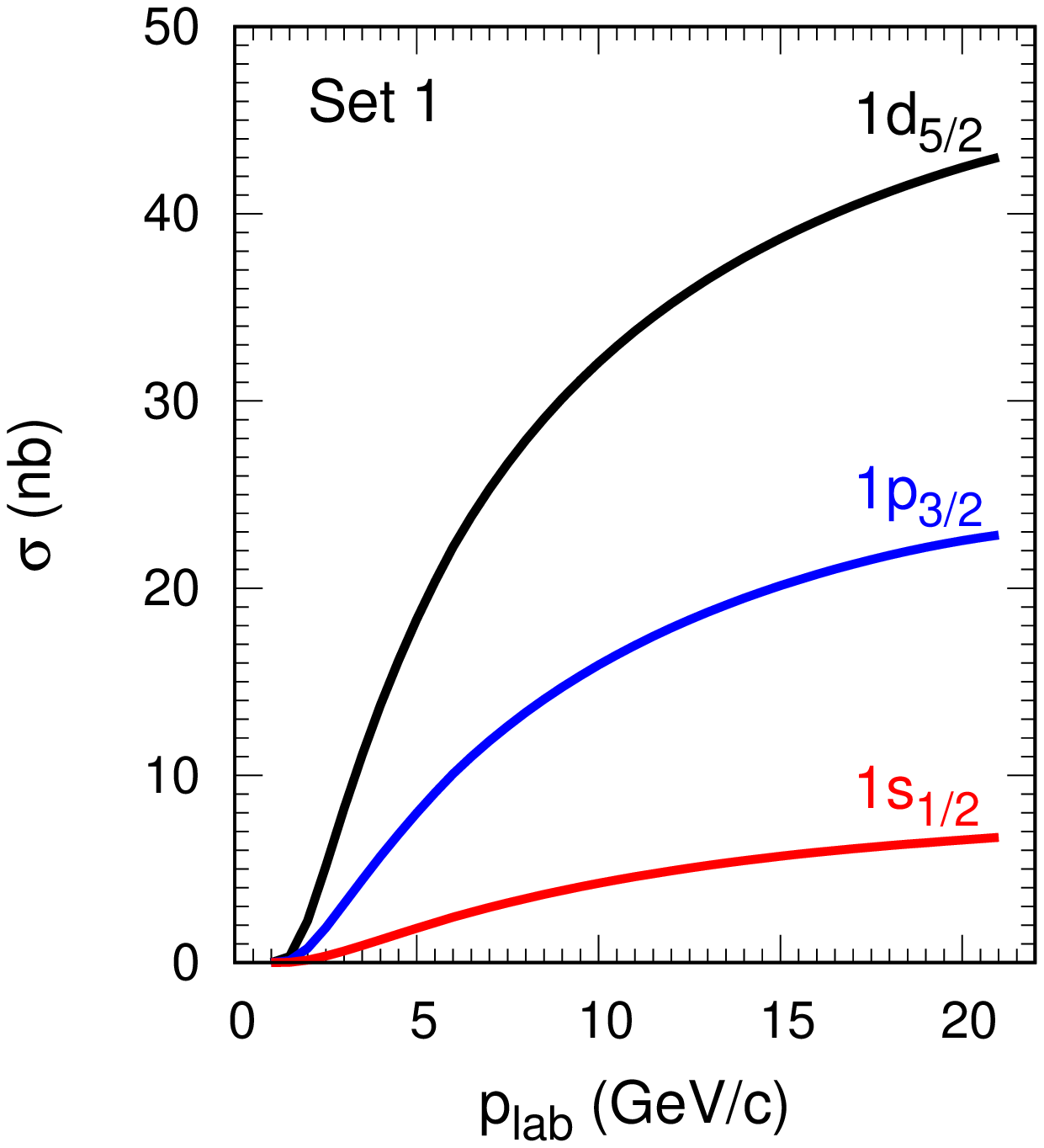}
     \includegraphics[scale = 0.5]{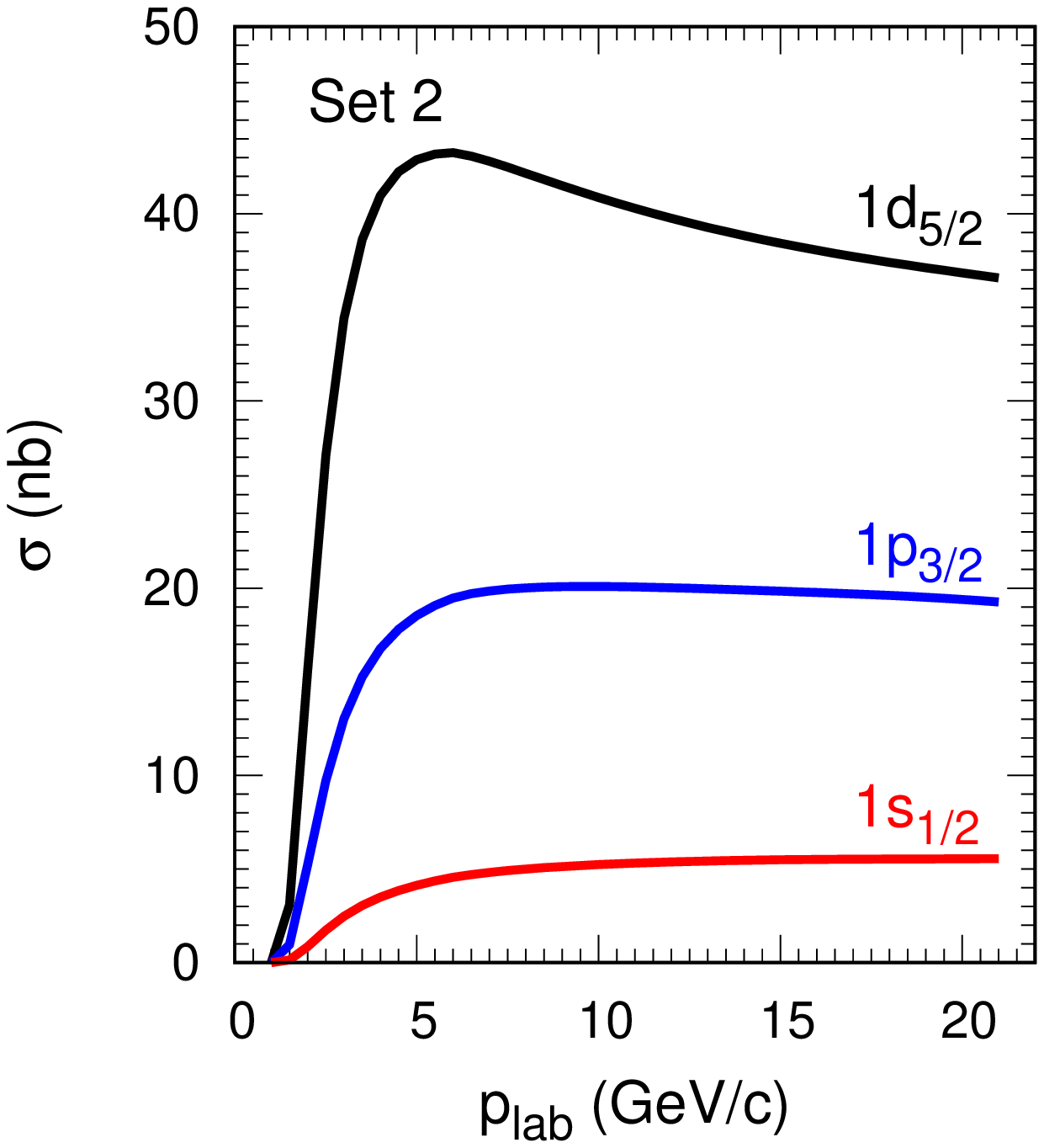}
\end{center}
\caption{\label{fig:md} Beam momentum dependence of the ${}^{40}_{~\Lambda}\mbox{Cl}$ hypernucleus production cross section
  in $\bar{p}\,{}^{40}\mbox{Ar}$ collisions for $\Lambda$ in $1s_{1/2}$, $1p_{3/2}$ and $1d_{5/2}$ states.
  The total cross sections  without (Set 1) and with (Set 2) $\kappa$ exchange are shown in the left and the right panel, respectively.}
\end{figure}
The robust signal of the $\kappa$ exchange is visible in the beam momentum dependence of the hypernucleus
production cross section shown in Fig.~\ref{fig:md}.
In calculations without $\kappa$ the $K^*$ exchange dominates and produces a smoothly growing cross section
with increasing beam momentum.
The $\kappa$ meson dominates at moderate beam momenta $\sim 1.5\div 3$ GeV/c (Fig.~\ref{fig:sigma_Lbar_L}).
This leads to the characteristic shoulder in $p_{\rm lab}$-dependence of the cross section of the hypernucleus production
and even to the appearance of the maximum for the $1d_{5/2}$ $\Lambda$ state.
We have checked that within the IA the maximum in the beam momentum dependence in the calculation including $\kappa$
becomes even more pronounced.
Hence this maximum is a clean manifestation of the $\kappa$ exchange and not an artifact of particular approximation
for the ISI/FSI effects.

\section{Summary and outlook}
\label{SumConcl}

In the present work, the calculations of the coherent hypernucleus production in $\bar pA$ collisions have been
performed. We have demonstrated that the pseudoscalar $K$ exchange is strongly suppressed for the bound $\Lambda$ states.
Thus the hypernucleus production is governed by the natural parity strange meson exchanges. Keeping only vector
$K^*$ exchange produces the smooth and structure-less increase of the cross sections of the hypernucleus production
with beam momentum. However, including the scalar $\kappa$ exchange leads
to the sharp change of the slope of the beam momentum dependence of the hypernucleus production cross sections
from increase to saturation between 4 and 6 GeV/c and even to the appearance of the pronounced maximum in the case
of $1d_{5/2}$ $\Lambda$ state. Hence we suggest that the measurement of the coherent $\Lambda$-hypernucleus production
cross section in $\bar pA$ collisions can be used as a test of possible exchange by the scalar $\pi K$ correlation.
These studies can be done at the planned \={P}ANDA experiment at FAIR and at J-PARC. Note that the antiproton
beam gives the unique opportunity to study $t$-channel meson exchanges, as, e.g. in the case of hypernuclear production
in $(\pi^+,K^+)$ reactions the process is governed by $s$-channel baryon resonance excitations \cite{Bender:2009cj}.  

With appropriate extensions the methods used here can be applied to the production of hyperons in low-energy unbound states,
thus allowing in principle to explore elastic scattering of $\Lambda$ and $\Sigma$ hyperons on nuclei. Experimentally,
such reactions could be identified e.g. by observation of the recoiling target fragment of mass number $A-1$. This may also help
to clarify further the still undecided question whether there are bound $\Sigma$ hypernuclei.

Of large interest is to establish the existence of the $\bar\Lambda$-hypernuclei which can be
produced in the process $(\bar p, \Lambda)$ with the capture of $\bar \Lambda$ in the residual nucleus.
However, that production process requires a large momentum
transfer to the struck proton (backward scattering). In this case, the description based on the reggeized $t$-channel meson exchange model
should be more appropriate for the elementary $\bar p p \to \bar \Lambda \Lambda$ amplitude. 

Charmed hadrons embedded in- or interacting with nuclei represent another related field of studies.
The coherent $\Lambda_c^+$-hypernuclei production in $(\bar p,\bar\Lambda_c^-)$ processes
has been explored theoretically  recently in Ref. \cite{Shyam:2016uxa}. The underlying $\bar p p \to \bar\Lambda_c^- \Lambda_c^+$
reaction on the bound proton has been described with a $t$-channel exchange by pseudoscalar $D^0$ and vector $D^{*0}$ \cite{Shyam:2014dia}.
Overall, the uncertainties in the charm sector are quite large due the lack of experimental data on the cross sections of the elementary
production processes and on nuclear FSI of the emitted charmed hadrons.
In particular, also the exchange by the scalar $D_0^*(2400)$ meson is possible in this case and can be distilled by using
the $\Lambda_c^+$-hypernuclei production in a similar mechanism.

\section*{Acknowledgements}
\label{Ack}

This work was supported by the Deutsche Forschungsgemeinschaft (DFG) under Grant No. Le439/9.
Stimulating discussions with M. Bleicher and M. Strikman are gratefully acknowledged.

\bibliographystyle{apsrev}
\bibliography{pbarHyp_plb}

\end{document}